\documentclass[11pt,a4paper]{article}
\usepackage{moriond,epsfig}

\def\Journal#1#2#3#4{{#1} {\bf #2}, #3 (#4)}

\def\PLB{{\em Phys. Lett.}  B}
\def\PRL{\em Phys. Rev. Lett.}
\def\PRD{{\em Phys. Rev.} D}

\def\Nature{\em Nature}
\def\ApJ{\em ApJ}
\def\ApJS{\em ApJS}
\def\AIPCP{\em AIP Conf. Proc.}
\def\MNRAS{\em Monthly Notices of the Royal Astr. Soc.}
\def\JCAP{\em Jour. of Cosmology and Astropart. Phys.}

\def\be{\begin{equation}}
\def\ee{\end{equation}}
\def\bea{\begin{eqnarray}}
\def\eea{\end{eqnarray}}

\newcommand{\lsim}{\lower 2pt \hbox{$\, \buildrel {\scriptstyle <}\over {\scriptstyle \sim}\,$}}

\begin{document}
\vspace*{4cm}
\title{THE QUEST FOR GRAVITY WAVE B-MODES}

\author{CLEMENT PRYKE}

\address{University of Minnesota Physics, 116 Church Street S.E.,\\
Minneapolis, MN, 55455, USA}

\maketitle
\abstracts{
One of the most exciting quests in all of contemporary science
is to find hints that in the first tiny fraction of a second after the
Big-Bang the Universe hyper-inflated by a factor of $\sim 10^{60}$.
Such inflation will have injected gravity waves into the fabric
of spacetime which will in turn have left a faint imprint in the
polarization pattern of the Cosmic Microwave Background.
This paper describes the history of polarization measurement,
the experimental optimization of this latest search for the
gravity wave imprint, and the current round of experiments
and their various approaches to the challenge.
}

\section{Introduction}

The study of the anisotropy of the Cosmic Microwave Background (CMB)
has taught us much about the origin, content and fate of the
Universe in which we find ourselves, and is one of the cornerstones
of what has come to be called the standard cosmological model (LCDM).
In this model the Universe began in a hot ``Big-Bang'' approximately
14 billion years ago and has been expanding and cooling ever since --- around
400,000 years after the beginning it made the transition from plasma (opaque)
to neutral gas (transparent) and the CMB is the highly re-shifted hot-object
(black-body) light which has been freely streaming through the
Universe ever since.
It therefore offers us a snapshot of the density pattern at this time
of ``last scattering''
on a spherical shell around the location where our galaxy would later
form.
Measurements of the total intensity ($T$) pattern by ground-based and
balloon-borne instruments had by 2003 resulted in stringent cosmological
constraints~\cite{tegmark03}, which were confirmed by the WMAP
spacecraft, and will shortly be further improved upon by the Planck space
mission.

Meanwhile efforts to observe the predicted polarization of the CMB
intensified and in 2002 the DASI experiment announced the
first detection~\cite{kovac02}.
The ``headless vector'' polarization pattern on the sphere can be broken
down into two scalar quantities dubbed $E$-modes and $B$-modes~\cite{zarsel97}.
If the electrons at last scattering are exposed to radiation fields
which have a quadrupolar component then Thompson scattering
results in a net polarization.
Flows of material generate such quadrupoles via Doppler
shifts along the direction of flow resulting in a ``pure gradient''
or $E$-mode pattern.
Since those flows are sourced by the density perturbations,
correlations between the $T$ and $E$ patterns also naturally result.
After last scattering the purity of the initial $E$-mode
pattern is slightly disrupted by small gravitational deflections
of the CMB photons as they travel to us through the forming
large scale structure.
This leads to a $B$-mode within the base model
which is referred to as the ``lensing $B$-mode''.

While the basic LCDM model is enormously successful it leaves
several fundamental questions unanswered.
From the earliest calculations it
was conventional to assume a flat power law initial perturbation
spectrum coming out of the Big-Bang~\cite{peebles70}.
There was no real reason to do so other than simplicity,
but, quite remarkably, recent CMB observations have proven it to be very
close to the truth!
Later the cosmo-genic theory known as
Inflation was invented~\cite{guth82} which naturally predicts
such an initial perturbation spectrum and also makes several
other predictions which have subsequently proven to be true,
such as the global flatness of space ($\Omega_{tot}=1$)
and the small degree of large scale anisotropy.

\begin{figure}[t]
\begin{center}
\resizebox{\textwidth}{!}{\includegraphics{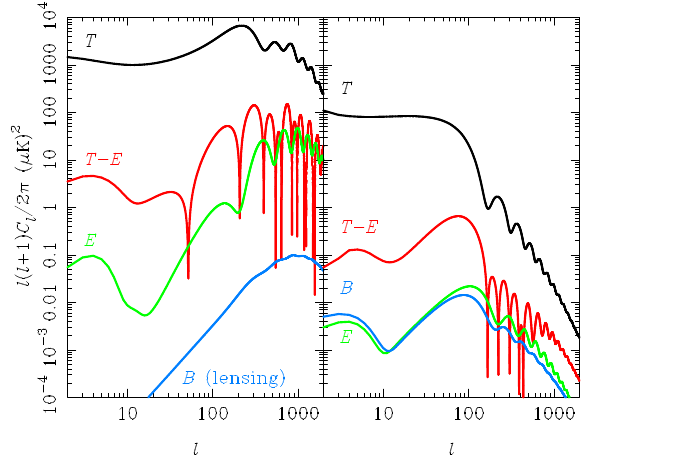}}
\end{center}
\caption{Theoretical CMB power spectra showing the
basic $T$, $E$ and $TE$ contributions from density perturbations
at left, along with the $B$-mode resulting from lensing
of that initial $E$-mode.
The tensor contribution from inflationary gravity waves
to each spectrum is shown at right for $r=0.2$.
We see that the inflationary $B$-mode at $\ell \lsim 100$ is potentially
detectable.
(Figure courtesy of A.\ Challinor)}
\label{fig:model-spectra}
\end{figure}

However Inflation is a radical theory positing expansion
of space by a vast factor ($\sim 10^{60}$) occurring at an incredibly
early time ($\sim 10^{-35}$~s), and hence at energies ($\sim 10^{16}$~GeV)
far, far above anything probed by terrestrial experiment.
Such radical ideas must be tested in all possible ways.
Fortunately inflation makes an additional prediction ---
it injects a background of gravitational waves (aka tensor perturbations)
into the Universe which
have been propagating through it ever since.
These distortions of the fabric of spacetime
will induce additional quadrupolar moments in the radiation field incident
on the electrons at last scattering which will in turn result in
both $E$-mode and $B$-mode polarization --- see 
figure~\ref{fig:model-spectra}.
The magnitude of the gravity wave component is conventionally
described by the tensor to scalar ratio denoted by $r$.

The motivations for making measurements of CMB polarization
can be summarized as follows:

\begin{itemize}
\item Measuring the $E$ and $TE$ spectra tests the basic
LCDM paradigm. At larger angular scales there is additional
information to be had regarding the process of reionization~\cite{kogut03}.
\item Measuring lensing $B$-modes at smaller angular scales can
provide information about structure formation, and in particular
neutrino mass.
\item Measuring $B$-modes at intermediate and large angular
scales holds out the tantalizing possibility of detecting Inflationary
gravity waves --- this has been referred to as the ``smoking
gun of inflation''.
\item The $TB$ and $EB$ spectra are identically zero in the standard
paradigm. However alternate Lorentz violating models might
produce a signal here~\cite{feng05}.
\end{itemize}

\section{Review of CMB Polarization Measurements to Date}

The first detection of CMB polarization was reported by
the DASI experiment in 2002~\cite{kovac02}.
Measuring over a broad range of angular scales
around $\ell \sim 200$ the $E$-mode was found to be consistent
with the LCDM expectation and inconsistent with zero at around the
$\sim 5 \sigma$ level, whereas the $B$-mode was consistent with
zero as expected.

In 2003 WMAP reported high significance detections of $TE$ correlation and 
the somewhat surprising result of a strong reionization contribution
at large angular scales~\cite{kogut03}, although this result was later
partially retracted.

\begin{figure}[h]
\begin{center}
\resizebox{\textwidth}{!}{\includegraphics{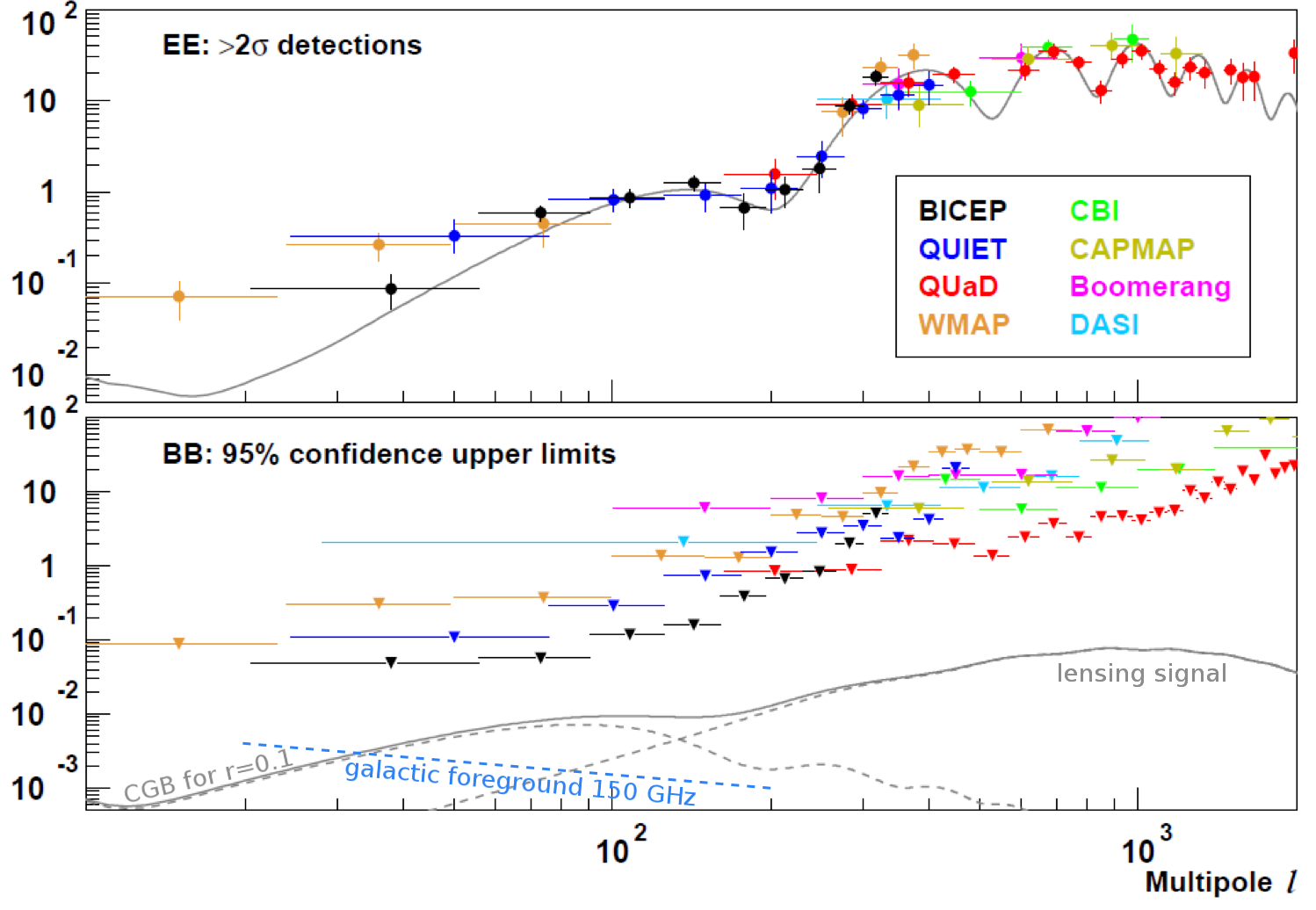}}
\end{center}
\vspace{-0.2in}
\caption{
Current results on CMB polarization.
The upper panel shows $E$-mode results compared to
the standard LCDM expectation.
The lower panel shows 95\% upper limits on $B$-modes, with the
dashed gray lines indicating B-mode signals
from lensing and a possible gravity wave signal with $r=0.1$; the solid gray line is the
sum of these two signals.
The dashed blue line indicates an estimate of the B-mode signal
from Galactic foregrounds (both synchrotron and dust) at 150 GHz
in a small, clean patch of sky.
(Adapted from Chiang et al.\ (2009) with the addition of more recent points from the
QUIET experiment and the foreground projections.)
}
\label{fig:currentres}
\end{figure}

To date $\sim 10$ experiments have reported detections of $E$
and $TE$ with only upper limits on $B$-modes to date ---
Figure~\ref{fig:currentres} shows the current situation.
At the moment the QUAD~\cite{brown09} and BICEP1~\cite{chiang09} experiments
lead the field at smaller and intermediate angular scales
respectively, with WMAP providing the only available information
at the largest scales.

\section{Optimizing the Quest for Gravity Wave $B$-modes}

The current best limit is $r<0.17$~\cite{keisler11}
from WMAP7+SPT temperature data~\footnote
{As we see in Figure~\ref{fig:model-spectra} gravity waves will add power
to all spectra including $T$.}.
However such $T$ based limits are now cosmic variance limited
and to go further we must search for $B$-modes.
As we see in figure~\ref{fig:model-spectra} there are two regions
where the detectability is maximized:
the ``recombination bump'' around $\ell \sim 80$ and the
``reionization bump'' at $\ell < 10$.
Clearly the recombination bump is potentially lost in
the rapidly rising lensing induced $B$-mode.
However to measure low multipoles one must use a large fraction
of the celestial sphere --- and much of it is obscured by foreground
emission from our own galaxy.
There is hence a complex trade-off between sky area
and observing frequencies when deciding how best to target this experimental goal.


\subsection{Galactic foregrounds and optimum observing frequency}

At low ($<30$~GHz) frequencies the sky brightness is dominated by
synchrotron radiation from relativistic electrons spiraling in
the magnetic fields of the galactic ISM (inter stellar medium).
At higher ($>300$~GHz) frequencies emission from galactic dust is dominant.
Since the dust is cold it is confined to a thin disk, whereas the high energy
electrons have a much larger scale height and fill the galactic halo.
It is a fortunate accident that the peak brightness of the $2.7$~K
CMB black body radiation at $\sim 150$~GHz is close to the minimum
of the total galactic emission.

\begin{figure}
\begin{center}
\resizebox{!}{3.5in}{\includegraphics{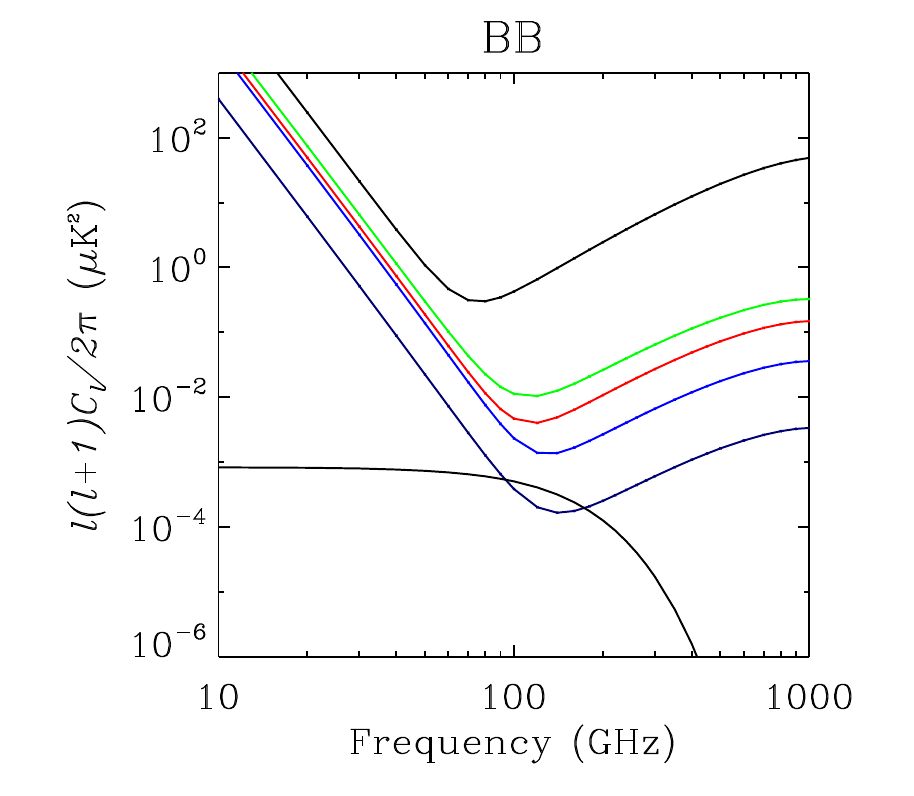}}
\end{center}
\vspace{-0.2in}
\caption{
The colored lines show the total projected polarized galactic emission
at angular scales of $80<\ell<120$ as a function of frequency
for sky coverage: from top to bottom, full, galactic latitude
greater than 10, 30, and 50 degrees,
and for a clean patch with radius of 10 degrees.
The lower black line shows the inflationary $B$-mode for $r=0.01$.
We see the asymmetry between dust and synchrotron foregrounds ---
(high frequency) dust emission falls much more rapidly with
increasing galactic latitude
causing the optimum frequency of observation to shift up
from $<100$~GHz towards $150$~GHz.
(Reproduced from Dunkley et al.\ (2008).)
}
\label{fig:foregrounds}
\end{figure}

It is very important to note that the maximum of the ratio of
CMB brightness to sync+dust is a function
of galactic latitude --- close to the disk one wants to go to lower frequencies
to get away from dust, whereas at high latitude the optimum frequency shifts up
as we see in Figure~\ref{fig:foregrounds}.
One must therefore be very careful of statements implying there is a single
best observing frequency ---
such statements always carry an implicit assumption as to the required sky area.

Moving from total intensity to polarization leads to additional complications
since the polarization fraction differs between synchrotron and dust, and with
position on the sky.
To connect the two requires (highly uncertain) modeling of the galactic
magnetic field --- see for example~\cite{odea12}.
The best current information comes from WMAP at low frequencies and
from extrapolations from IRAS at high frequencies.
Figure~\ref{fig:foregrounds} shows a projection ---
information from Planck will shortly massively increase our knowledge
of polarized foregrounds.

\subsection{Small patch or big patch?}

If one decides to target the reionization bump there is no
choice as to sky coverage --- to resolve modes at $\ell<10$ one
needs as large a fraction of the sky as possible.
Since we already know $r \lsim 0.2$ this means that heavy
foreground ``cleaning'' will be required to reach the cosmological
signal.
This dictates making measurements at several frequencies and
decomposing them as sums of multiple components ---
a well established technique in WMAP (and now Planck) analysis.
The ultimate limitations of cleaning are hard to predict but the
prospect of pulling out ``ghost-like'' $\ell < 10$ cosmological $B$-modes
which are submerged by many orders of magnitude beneath galactic
foreground is daunting to say the least.
Nevertheless the projected ability to set limits of $r \lsim 0.03$ with
Planck has been claimed~\cite{efstathiou09}.
Measuring close to the whole sky at many frequencies is something
that many would claim is only possible in
space~\footnote
{But see the PIPER and CLASS experiments in Section~\ref{sec:expt}.}.

If one instead chooses to target the recombination bump there is a huge
benefit --- we can focus on small patches of sky at high galactic latitude
which have foreground contamination at least an
order of magnitude lower.
From a simplistic signal-to-noise point of view when aiming for an initial
detection one wants to
concentrate the available sensitivity onto as small a patch
of sky as possible.
The lower limit is then set by the need for the patch to be
at least as big as the largest angular scale of interest.
Since the $E$ and $B$ mode patterns are non-local in practice
one needs a patch several times larger than this to maintain
separability~\cite{agnesferte}.
In practice when targeting the $\ell \sim 80$ recombination bump
a patch of area $\sim 500$ square degrees is appropriate.

\subsection{Angular resolution}

Once one has decided the observing frequency, and patch of sky
to be observed, the remaining variable is the size of the telescope
aperture --- the angular resolution.
In principle the requirement is modest --- the angular scale
corresponding to $\ell \sim 80$ is adequately resolved by
a 25~cm aperture at 150~GHz.

However if higher angular resolution and sufficient sensitivity
are available then the lensing contribution at larger angular
scales can be computed and subtracted in a map sense ---
the lensing component can be cleaned out.
From Figure~\ref{fig:currentres} we see that this
becomes necessary around $r \sim 0.01$.
In addition, of course, measuring the lensing $B$-modes is an
important goal in of itself allowing several scientific
results including constraints on the neutrino mass.

\subsection{Detector Technologies}

The two main detector technologies used for CMB detection
are bolometers (photons go to heat) and coherent amplifiers
(photons go to voltage on a wire).
Either of these can be used in direct imaging (telescope beam
scans around on the sky) or interferometric systems (multiple
telescopes ``stare'' at the sky and outputs are cross correlated).
The DASI and CBI experiments were 30~GHz interferometers based on
(coherent) HEMT amplifiers.
The WMAP, CAPMAP and QUIET experiments were direct imaging HEMT based
systems.
The other experiments shown in Figure~\ref{fig:currentres} (Boomerang,
QUaD and BICEP) were direct imaging bolometer systems.
To date bolometer experiments have held the edge in terms
of per detector sensitivity.

\subsection{Polarization Modulation and Beam Systematics}

To measure linear polarization anisotropy in a direct imaging experiment
one needs to measure brightness differences scanning across the sky
with detectors sensitive to different polarization directions.
Even from the best sites on the ground one is looking at the CMB
through a ``glowing screen'' whose brightness variations
are many times the size of the CMB anisotropies (think clouds).
This problem is mostly alleviated on balloons or in space,
but all detectors are subject to intrinsic fluctuations
($1/f$ noise) at low frequencies which have a similar
practical effect.
However the atmospheric emission is largely unpolarized
and $1/f$ noise largely common mode between co-located detectors.
Therefore one can overcome these effects by either fast modulation of
the polarization sensitivity direction of a single detector
(faster than the $1/f$ and scan rate), or by simultaneous differencing
of orthogonal pairs of detectors (combined with later re-observation
at another angle with the same, or another, pair of detectors)~\footnote
{Pair differencing is popular --- of all the bolometer
experiments mentioned in Table~\ref{tab:exptsum} EBEX and PIPER
may be the only ones which do not have co-located pairs of detectors.}.

When modulating the modulator must not change the beam position
and shape on the sky.
When pair-differencing each half of the pair must have the
same beam position and shape on the sky.
Inasmuch as these conditions are violated then leakage from
the much stronger $T$ and $E$-mode spectra will occur
into the $B$-mode spectrum.

In general beam imperfections are a fraction of the beam width
so they are much more critical for the big beam (small aperture)
experiments.
However many of the effects at least partially cancel by
rotating the telescope with respect to the sky and
repeating the observations.
Such rotation may take place due to the rotation of the Earth
(at non-polar sites) and/or by including an additional
line-of-sight (LOS) axis into the telescope mount.
For some pair differencing experiments a major
problem is ``A/B centroid mismatch'' where the beams of the orthogonal
pair are offset from one another on the sky.
This causes leakage from the gradient of $T$ into polarization.
However consider rotating by 180~degrees --- now the sense
of the leakage is reversed and under co-addition it cancels.

Some beam imperfections do not cancel under rotation
of the telescope.
For this reason some pair-differencing experiments
also incorporate a slow (stepped) polarization modulator.
So long as the behavior of the upstream optics is
polarization independent (a property always required
of a modulator) then one can achieve cancellation
of beam systematics by repeating the observations
at a set of discrete modulation angles.

Analysis mitigation is also possible ---
if one has knowledge of the beam imperfections and a
$T$ map (from another or the same experiment) one
can compute the leakage and subtract it out.
Going further one can instead ``project out'' the
potentially contaminated modes and avoid having to know
the beam imperfections a priori.
Many frequently referenced estimates of the needed degree
of beam performance ignore such observation and
analysis based mitigation and hence conclude that
much higher beam purity is needed than is in
fact the case~\cite{huhedmanzaldariaga03}.

\subsection{Ground/Balloon/Space}

Observing from the ground modern detectors are ``background limited''
meaning that photon arrival statistics of the atmospheric
emission dominate the statistical noise.
Thus when observing from above the Earth's atmosphere
dramatically better per detector performance can be achieved.
However even ``long duration'' balloon flights are currently
$\lsim 10$ days whereas ground based experiments can and do
observe for up to $\sim 1000$ days.
Historically both techniques have delivered final results
of comparable quality when equipped with similar numbers
of detectors.

Observing from space is a different matter --- now one has the
low noise level \emph{and} years of exposure.
However note that while in principle a CMB spacecraft could
concentrate this awesome sensitivity on a small patch of
clean sky, in practice --- because they have the unique
capability to observe the whole sky --- they always do.
Therefore the deepest small patch maps at any given time are always those
from ground/balloon experiments.

\section{Current/Future Experimental Efforts}
\label{sec:expt}

There are a large number of experimental
efforts focused on CMB polarization.
Table~\ref{tab:exptsum} summarizes these with some
additional information on each given below.

\subsection{Space and Balloon}

\hspace{0.25in} {\bf Planck} is a 1.5~m aperture ESA space mission that launched in
2009.
The High Frequency Instrument used bolometer detectors
similar to those used in the Boomerang, BICEP1 and QUaD experiments,
and completed operation in early 2012.
The Low Frequency Instrument uses HEMT amplifiers and is still operating.
Planck makes full sky maps and improves over WMAP in terms of
frequency coverage (9 versus 5 channels), angular resolution ($\sim 3 \times$)
and sensitivity ($\sim 10 \times$).
Initial CMB results are expected in early 2013 with polarization results
not expected until early 2014.

{\bf EBEX} is a 1.5~m aperture balloon based experiment funded by
NASA which has been under preparation for several years.
Its first science flight will be a $\sim 10$~day circumpolar
(``long duration'' or LD) flight in Anarctica in late 2012.
EBEX has 1500 TES bolometer detectors modulated by a continuously
spinning half wave plate at 150, 250 and 410~GHz.

{\bf SPIDER} is a balloon based array of 0.3~m aperture telescopes
funded by NASA.
The telescope optics and focal planes are very similar to
BICEP2 and Keck-Array.
A circumpolar flight is planned for late 2013 with focal
planes at 90, 150 and 280~GHz and a total of $\sim 2000$
TES detectors.

{\bf PIPER} is a third NASA funded balloon program emphasizing
high frequencies (200, 270, 350, 600~GHz) with smaller
resolution ($\sim 0.3$~meter) apertures and large numbers
of TES detectors (5000 per frequency).
Only one frequencies will be active in any given flight
and eight ``standard duration'' (SD) flights are planned split between northern
and southern hemispheres.
The goal is to cover a large fraction of the sky and target the
reionization bump.

\subsection{Ground Based in Chile}

All the following experiments are (largely) funded by
the US NSF and located at various sites in the Atacama
desert in Chile.

{\bf QUIET} is 1.4~m aperture telescope utilizing HEMT
amplifiers.
It has released 43~GHz results (see Figure~\ref{fig:currentres}),
and 90~GHz results are expected soon.
It is currently unclear if the QUIET program will continue.

{\bf POLARBEAR} is a 2.5~m aperture telescope.
Deployment was completed in early 2012 and the focal plane
is equipped with $\sim 1300$~TES detectors operating
at 150~GHz.

{\bf ABS} is a 0.3~m all cold reflecting telescope.
It has $\sim 500$ TES detectors at 150~GHz and has been running
since early 2012.
ABS has a warm waveplate outside the cryostat which spins at 2.5~Hz.

{\bf ACTpol} is a polarized receiver for the existing
6~m ACT telescope which will deploy in late 2012.
ACTpol is \emph{not} emphasizing inflationary $B$-mode
detection.

{\bf CLASS} is unique amongst ground based
experiments in that it is targeting the reionization bump.
It will observe at 40, 90 and 150~GHz and deployment is planned
to start in 2013.

\subsection{Ground Based at South Pole}

The following experiments are (largely) funded
by the US NSF and located at the South Pole station
in Antarctica.

{\bf SPTpol} is a polarized receiver for the existing
10~m SPT telescope which deployed in early 2012
and is now operating.
It has 1500 total TES detectors at 100 and 150~GHz.

\begin{figure}
\begin{center}
\includegraphics[width=\textwidth]{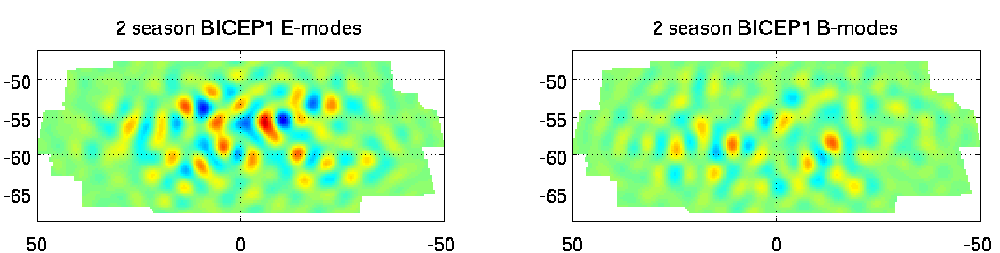}
\includegraphics[width=\textwidth]{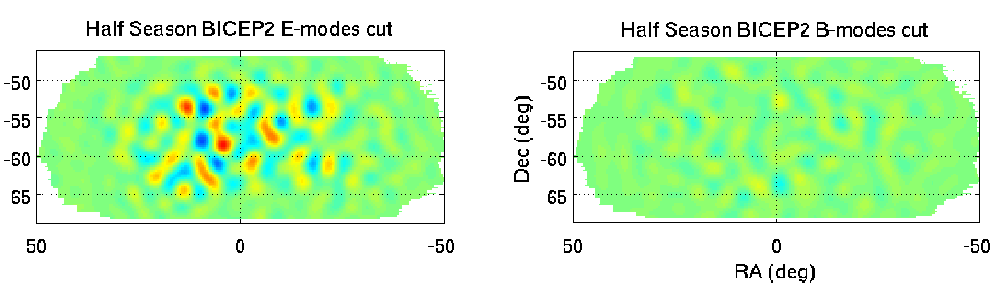}
\end{center}
\vspace{-0.2in}
\caption{Maps of the CMB polarization $E$-mode and $B$-mode filtered to the angular
scale range where the inflationary gravity wave signal is expected to
be most prominent ($50<\ell<120$).
The upper maps are the published BICEP1 results which hold the current world record
in terms of sensitivity ($r<0.72$, Chiang et al.\ (2010)).
The lower maps are new preliminary BICEP2 results using half a season of data.
The $E$-mode maps were already signal dominated for BICEP1 and so show little change
going to BICEP2.
However the BICEP2 $B$-mode maps show a large decrease in noise due to
the much higher sensitivity of the BICEP2 instrument.}
\vspace{-0.15in}
\label{fig:b1_vs_b2_eb_maps}
\end{figure}

{\bf BICEP2} is a 0.3~m refracting telescope
with 500 TES detectors at 150~GHz.
As of mid 2012 it is midway through its third and final
season of observation.
The proven on sky sensitivity of BICEP2 is $\sim 10 \times$ higher
than that of BICEP1 --- comparable to that of the Planck
spacecraft but concentrated on a small patch of ultra
clean sky.
Figure~\ref{fig:b1_vs_b2_eb_maps} shows $E$ and $B$-mode maps
from BICEP2.

{\bf Keck-Array} (aka SPUD) is an array of BICEP2 like
telescopes.
It operated through the 2011 season with three receivers
at 150~GHz, and is operating for 2012 with five such receivers.
For future seasons some receivers may be reconfigured
to 90 and/or 220~GHz.

\begin{table}[t]
\caption{Parameters of current and upcoming CMB telescopes.\label{tab:exptsum}}
\vspace{0.4cm}
\begin{center}
\begin{tabular}{|c|c|c|c|c|c|c|}
\hline
Name & Type & Deploy/Fly & Aperture & Freq.\ (GHz) & Detectors & Modulation$^a$ \\
\hline
Planck & Space & 2009 & 1.5~m & 30--860 &  & PD/LOS \\
EBEX & LD Balloon & late 2012 & 1.5~m & 150/250/410 & 1500 TES & SR/FM \\
SPIDER & LD Balloon & late 2013 & 0.3~m & 90/150/280 & 2000 TES & SR/PD/SM \\
PIPER & SD Balloon & 9/2013 & $\sim 0.3$~m & 200/270/350/600 & 5000 TES & SR/FM \\
QUIET-I & Chile & 2008--10 & 1.4~m & 40/90 & 19/90 HEMT & SR/FM/LOS \\
\tiny{POLARBEAR} & Chile & early 2012 & 2.5~m & 150 & 1300 TES & SR/PD/SM \\
ABS & Chile & early 2012 & 0.3~m & 150 & 500 TES & SR/PD/FM \\
ACTpol & Chile & late 2012 & 6~m & 90/150 & 3000 TES & SR/PD \\
CLASS & Chile & late 2013 & ? & 40/90/150 & ? & SR/? \\
SPTpol & South Pole & early 2012 & 10~m & 100/150 & 1500 TES & PD \\
BICEP2 & South Pole & early 2010 & 0.3~m & 150 & 500 TES & PD/LOS \\
Keck-Array & South Pole & early 2011 & 0.3~m & 150 & 2500 TES & PD/LOS \\
\hline
\end{tabular}
\end{center}
\vskip 5pt
$^a$ SR = sky-rotation, FM/SM = fast/slow modulator (HWP, VPM or phase-switch),
PD = pair-difference, LOS = line-of-sight rotation of whole telescope.
\end{table}

\subsection{Future Space Mission?}

The 2007 Bpol proposal to ESA consisted of an array
of small refracting telescopes and was not selected.
The US 2010 Decadal Survey recommended that NASA reconsider
a future CMB space mission mid-decade.
In 2010 the European community tried again with the more ambitious
1.2~m 6000 detector COrE mission which incorporated a large
reflective polarization modulator.
This was rated just below the selection cutoff for
further study.

\section{Conclusion}

Inflation is often talked about as a part of the ``standard
cosmological model''.
It is perhaps more correct to say that this audacious theory
is compatible with observations --- it naturally produces the
sort of Universe in which we find ourselves:
spatially flat, isotropic and with scale-free, adiabatic initial
conditions.
It is therefore of critical importance to search for the
final, and so far unobserved, prediction of Inflation ---
a background of primordial gravitational waves.
If such waves exist at the time when the CMB photons
last scatter they will imprint a very specific $B$-mode
polarization pattern.

This paper has described the optimizations and trade-offs
necessary to search
for the gravitational wave signal in terms of frequency,
sky-coverage, detector technology, modulation and experiment
location.
The current experimental ``landscape'' has also been summarized:
many groups are pushing hard to detect this possible signal
using a range of approaches, and emphasizing different parts
of the problem.
If nature is kind, and $r \geq 0.02$ then we should detect it soon!

\section*{Acknowledgments}

The author wishes to thank his collaborators and many other
members of the CMB community for useful discussions,
the NSF (grant number ANT-1110087) for funding, and the conference
organizers.

\end{document}